\documentclass[aps,pre,twocolumn,preprintnumbers,showpacs,superscriptaddress]{revtex4}

\usepackage[final]{graphicx}
\usepackage{dcolumn}
\usepackage{bm}
\usepackage{floatflt,epsfig}
\usepackage{float}
\usepackage{amsmath}
\usepackage{amsfonts}
\usepackage{amssymb}
\usepackage{mathrsfs}
\usepackage[latin1]{inputenc}

\DeclareTextSymbol{\degre}{T1}{6}
\DeclareTextSymbol{\degre}{OT1}{23}

\DeclareMathOperator{\EE}{\mathscr{E}}

\DeclareMathOperator{\CC}{\mathscr{C}}
\DeclareMathOperator{\DD}{\mathscr{D}}

\DeclareMathOperator{\TT}{\mathscr{T}}

\begin{document}
\title{Transient optical response of ultrafast nonequilibrium excited metals: Effects of electron-electron contribution to collisional absorption}
\author{J.P. Colombier}
\affiliation{Laboratoire Hubert Curien, Universit\'e Jean Monnet, UMR CNRS 5516, 18 rue Benoît Lauras, 42000 Saint-Etienne, France}
\author{P. Combis}
\affiliation{CEA/DAM Ile de France, Dept de Physique Th\'eorique et Appliqu\'ee, 91297 Arpajon, France}
\author{E. Audouard}
\affiliation{Laboratoire Hubert Curien, Universit\'e Jean Monnet, UMR CNRS 5516, 18 rue Benoît Lauras, 42000 Saint-Etienne, France}
\author{R. Stoian}
\affiliation{Laboratoire Hubert Curien, Universit\'e Jean Monnet, UMR CNRS 5516, 18 rue Benoît Lauras, 42000 Saint-Etienne, France}%
\date{\today}

\begin{abstract}
Approaching energy coupling in laser irradiated metals, we point
out the role of electron-electron collision as an efficient
control factor for ultrafast optical absorption. The high degree of laser induced electron-ion nonequilibrium drives a complex absorption pattern with consequences on the transient optical properties. Consequently, high electronic temperatures determines largely the collision frequency and establish a transition between absorptive regimes in solid and plasma phases. In particular, taking into account umklapp electron-electron collisions, we performed hydrodynamic simulation
of laser-matter interaction to calculate laser energy deposition
during the electron-ion nonequilibrium stage and subsequent
matter transformation phases. We observe strong correlations between optical and thermodynamic properties according to experimental situations. A suitable connection between solid and plasma regimes is chosen in
accordance to models that describe behavior in extreme, asymptotic regimes. The proposed approach describes as well
situations encountered in pump-probe type of experiments, where the state of matter is probed after initial excitation. Comparison with experimental measurements shows simulation results which are
sufficiently accurate to interpret the observed material behavior. 
A numerical probe is proposed to analyse transient optical
properties of matter exposed to ultrashort pulsed laser
irradiation at moderate and high intensities. Various thermodynamic states are assigned to the observed optical variation. Qualitative indications on the amount of energy coupled in the irradiated targets are obtained. 
\end{abstract}


\maketitle
\section{Introduction}

Laser-induced phase transitions are extensively investigated by
means of optical measurements due to their fundamental interest
and potential in material sciences and engineering. Optical
probing was identified as a relatively straightforward way to
time-resolve changes of material properties. However, singular
effects in the material response in extreme conditions are not always easy to identify
in the sequence of matter alterations \cite{Sun94,
Groeneveld90,Guo00,Norris03,Widmann04}. Although more accurate measurements using X-ray or electron diffraction have emerged to probe matter transformation with atomic resolution~\cite{Siders99,Siwick03}, optical probing is still a powerful technique to resolve evolution changes in material transformations in a wide range of thermodynamical states. First investigations of
ultrashort-pulsed laser energy relaxation in metals concentrated on a wide range of effects including pump-probe reflection and transmission efficiencies \cite{Eesley86}, time-resolved particle emission and two-photon photoemission experiments \cite{Fann92}. Subpicosecond lasers have been used to resolve
nonequilibrium heating in Au and Cu \cite{ElsayedAli87,Schoenlein87} and to measure the
electron-phonon coupling for a variety of metals
\cite{Brorson90,Sherman89}. Modelling the correct behavior of the material
properties under strong excitation is still a
challenge for understanding transport properties and relaxation
processes in nonequilibrium systems
\cite{Bejan97,Gusev98,Rethfeld02}. Irradiated by an intense laser
beam, a material undergoes a complex series of transformations,
from solid-solid, solid-liquid, liquid-gas, up to the plasma phase. Each transformation has a particular optical signature and
the description of the evolution remains a fundamental issue
\cite{Celliers93,Ng98,Komashko99,Eidmann00}. Hence, comparison of
experiments with dedicated models is needed to validate the
modelling approach even though experiments on extreme conditions
of matter are difficult to perform and the data are scarce and often
contradictory. Nevertheless, since experiments involve complex
interconnected phenomena, various models with different
degrees of complexity are required to decode the experimental information.

A specificity of ultrafast laser excitation is its highly transient nature. Ultrafast laser irradiation can push matter in extreme high pressure, high temperature, nonequilibrium states where hot electrons interact with colder
ions \cite{Anisimov74}. Hereby, nonequilibrium describes the
temperature mismatch between hot electrons and ions, while each
interacting subsystem is considered thermalized on the timescales
involved. If this is a crude approximation in the first moments of interaction, the establishment of a high collision rate will reduce the domain of incertitude. Subsequently, the properties of the material are
governed by the relaxation paths towards the equilibrium state
\cite{Kaganov57}. Moreover, since optical probing only involves the electronic system,
information on lattice structural changes must be inferred
indirectly from changes in the electronic system. From a
fundamental and theoretical point of view, interesting features
emerge when the pulse width becomes shorter than the
electron-phonon relaxation time scale for energy deposition into the
solid \cite{Anisimov02}. For this reason, we will be approaching
systems whose dynamical evolution exhibits a pattern driven by the
electron-electron and electron-ion collisional interactions (such as reflectivity or absorption coefficients). The
energy distribution among each species determines changes within the
excited material which affect the relative transport properties.
Since optical properties depend strongly on the ionization degree
which is in turn controlled by temperature and density evolution,
a hydrodynamic approach is required. A coupled electromagnetic and
hydrodynamic formalism, associated to specific optical and thermal models reflecting
nonequilibrium features, provides an effective way to reproduce
laser-matter experiments. Such simulations have been performed using
the ESTHER code, developed by the Commissariat
à l'Energie Atomique, France. Details to the code assumptions and approaches are given in Ref.[\onlinecite{Colombier05},\onlinecite{Colombier06}], and will be briefly reviewed in the next sections, focused on absorption and thermodynamic description.

A full theoretical description of solid-laser interaction requires the selfconsistent treatment of hydrodynamics and the evolution of the transport properties which are associated. Despite many efforts, a global view including the multitude of interaction and relaxation processes is still incomplete and only particular aspects are understood at this time.  The analysis of the transient optical properties of a laser-induced dense nonequilibrium plasmas requires a description of the electron degenerecency effects. Different
approaches have been used to model interaction mechanisms and
subsequent energy absorption. For considering the role of these interactions
on degenerate matter properties, ionic and electronic structure
effects in relation to transport properties have to be calculated 
\cite{Lee84,Perrot87,Desjarlais02,Faussurier06}. Some of the
studies dealing with ultrafast absorption phenomena are focused on the complex
effect of simultaneous contribution of collision processes on the
kinetic equation \cite{Lugovskoy94,Lugovskoy99,Rethfeld02}. In considering the nonequilibrium dynamics of
electrons in metals, the mutual influence of \emph{(e-e)} and
\emph{(e-p)} interactions remains an issue of interest. Although quantum mechanical approaches and kinetic treatments have been developed to caracterize the absorption rate in dense plasmas and solids \cite{Silin81,Bornath01,Schulz95} and significant progress was made, the formulation of the electrons excitation through all these regimes remains difficult and limited. Consequently, we concentrate our calculations on the effects resulting from bringing together basic models to discriminate dominant mechanisms in several experimental situations, encompassing a large variety of thermodynamic states. For excitation below certain relaxation times (electron-phonon equilibration), nonequilibrium
appears, particularly in the femtosecond excitation regime. The dominant effects on
the nonlinear optical dynamics arrive from the Pauli exclusion
principle and the Coulomb electron-electron interactions among the free carriers \emph{(e-e)}.
The phonons also play an important role, especially on
picosecond time scales. During and following their excitation, the
electrons undergo a number of different relaxation stages as they
scatter among themselves via the Coulomb interaction and with the
phonons via the electron-phonon \emph{(e-p)} interaction. In optical absorption, contribution may appear from electronic processes that allows momentum variation and conductivity changes. The \emph{(e-e)} interaction contributes indirectly to the inverse bremsstrahlung process of absorption since it screens the electron-ion pseudo-potential, and directly via umklapp processes. 
For simplicity and according to some other investigations dedicated to optical probing
\cite{Huettner94,Bejan97,Fisher01}, we adopt a phenomenological
approach as being well suited for integration in hydrodynamic codes. In this natural way, collision rates are described under the
relaxation-time approximation for both \emph{(e-e)} and
\emph{(e-p)} \cite{Ashcroft}. Obviously, this simplified approach
can not describe specific features of the collision process, but, along side with different kind of
experiments, it is adequate to improve \emph{ad hoc} transport
models in solid material up to the plasma phase. This will help reproducing
transient evolution of the optical properties. By measuring the
amplitude of the optical field emerging from the irradiated
sample upon reflection, one can gain valuable insight into the interplay between
many-body interactions and quantum confinement effects during very short time
scales. 

The aim of this paper is to
provide a deeper insight into the transport characteristics and their
consequence on the optical properties in nonequilibrium systems,
such as ultrashort laser-excited metals. With support of numerical calculations, we concentrate our investigations on several issues. First of all, the effect of electron-electron collision on the transient optical properties is modelled. Our original motivation was to insert a more accurate description of the laser absorption in the ESTHER code. In order to validate our modelling assumptions, we perform simulations of reflectivity dependence on the
angle of incidence, polarization and intensity while comparing the results with experimental data
available in the literature \cite{Fedosejevs90,Milchberg88}. Our
results reveal that calculations match the experimental data with
great accuracy when considering the effect of \emph{(e-e)}
collision rate enhancement during laser irradiation. To extend our investigations, a great challenge is to compare the time evolution of the calculated optical properties with time-resolved experiments. In fact, with subpicosecond laser pulses, it is also possible to probe in a
time-resolved fashion the charge carrier dynamics in metals by a
pump-probe optical system. Recent measurements on reflectivity
and transmittivity evolution under ultrafast laser irradiation
in pump-probe type of experiments have been reported \cite{Widmann04}. To realize a similar numerical pump-probe experiment, further developments are required to identify potentially-induced phenomena in
the framework of an optical, thermal and hydrodynamical study. Straightforward numerical modeling of such problems is a difficult challenge due to the various nature of underlying physical phenomena. While performing an analysis of the measured optical properties, we propose an interpretation of the results  by splitting the behavior into two stages : a dense and nonequilibrium plasma
phase following by a hot expanded plasma regime. Our laser-matter numerical
tool demonstrates new abilities to investigate nonequilibrium and associated collisional effects on the material
states ranging from cold solid to hot plasma. 

The rest of this
paper is organized as follows. We firstly detail the modelling
of the laser absorption by the material in Secs. II and III. In
Sec. IV, we make use of comparison with available experimental
data sets to validate our numerical modelling. We then broaden the
range of simulation by investigating the electromagnetic interactions relative to pump-probe numerical experiments. These experiments ranging from high intensities to below ablation threshold excitation, furnish arguments that demonstrate the applicability of
such a numerical tool. Conclusions are drawn in the last
section.

\section{Electromagnetic excitation by ultrashort laser pulse}
\subsection{Laser-matter interaction background}
When electromagnetic irradiation is shorter than several ps, the equilibration time between the
solid lattice and the laser-heated electrons is longer
than the laser pulse length. Thus, thermal equilibrium could not be
ensured and the thermal state of the solid lattice is not well
known. In order to understand the complexity of material
absorption under the laser irradiation, the modelling of
laser-matter interactions requires the interconnection of
different physics fields, including optics, heat transfer, fluid
dynamics, and phase transitions in nonequilibrium frames. Among
different models that treat the interaction between ultrafast
laser light and metallic targets at moderate intensities and
provide scenarios of material
ejection \cite{Rethfeld02,Ki00,Schafer02,Itina04,Lorazo06}, the
hydrodynamic approach employing tabulated
Bushman-Lomonosov-Fortov~\cite{Bushman93} multiphase equations of
state (EOS) addresses reasonably well the
complexity of the problem. It also has the potential to reveal
plausible evolution paths for the excited matter that have also
relevance for improving the laser interaction
process \cite{Colombier06,Colombier07}. The 1D hydrodynamic code ESTHER
includes a Two-Temperatures Model to reproduce electronic thermal
diffusion which takes place during electron-ion nonequilibrium. It
is extended by a nonequilibrium hydrodynamic model decoupling
electron and ion pressure components which both contribute to the
ultrafast matter deformation according to the Euler equations.
Fluid equations are solved to account for the conservation of energy,
momentum, and mass in a Lagrangian formalism. The specificity of
this approach is the flexible dimension of the cells which deform
together with the material and permit an accurate description of
shock propagation. The simulation accounts for mechanical
processes and particularly the effects on material strength that
occur in shock propagation. This is due to constitutive models which
describe the pressure and temperature dependence of shear modulus
and yield strength \cite{Steinberg80}. This code is enhanced with a wave-equation
solver to calculate the electromagnetic field in the material
during laser-matter interaction. To determine optical response,
especially in the presence of a density gradient, Maxwell equations are reduced to the Helmholtz equation by using
slowly-varying-envelope approximation for the electric field. This
is justified because, for a laser pulse of $\tau=150$ fs, the
wavelength, $\lambda=800$ nm, is smaller than the coherence length
$L_{\tau}=c\tau$, where $c$ is the speed of light.

\subsection{Optical property modelling}
If the idealized density step-profile is no longer actual, the Fresnel equations cannot be used and it becomes necessary to calculate the absorption
in a non-uniform density profile. The electric field is calculated
numerically by solving the wave equation, using a one-dimensional
mesh. The cell coordinates evolve in time due to the Lagrangian
approach used to solve the hydrodynamic equations. For two pulses excitation, the contribution of each pulse is accounted for. 
The subsript $\alpha$ refers to the pump laser field at the frequency
$\omega_{\alpha}$ and $\beta$ refers to the probe one at the
frequency $\omega_{\beta}$. We consider a planar wave propagating along the $z$-axis.  We denote $\vartheta$ the angle which the normal to the wave makes with the $z$-axis. Each component ($x,y,z$) of the electric field is defined by the equation $\tilde\EE^{x,y,z}_{\alpha,\beta}=\tilde{A}^{x,y,z}_{\alpha,\beta}(z) \exp(i k^{0}_{\alpha,\beta} x \sin \theta)$ where $k^{0}_{\alpha,\beta}=\omega_{\alpha,\beta}/c$ is the vacuum wave number. We write the following Helmholtz equation which is solved two times for the complex amplitudes of
each components ($\alpha,\beta$) of the electric fields:
\begin{equation}\label{three}
\frac{\partial^{2} \tilde A^{x,y,z}_{\alpha,\beta}}{\partial z^{2}}
+\tilde k_{\alpha,\beta}^{2} \tilde
A^{x,y,z}_{\alpha,\beta}=0.
\end{equation} 
Here, the complex vector $\tilde k$ depends on the dielectric function $\epsilon$:
\begin{eqnarray} 
\tilde k_{\alpha,\beta}^{2}=\left(k^{0}_{\alpha,\beta}\right)^{2}[\epsilon-\sin^{2} \vartheta] \label{k0}.
\end{eqnarray}
In the solid case, the complex vector $\tilde k$ is
 defined as a function of the refractive index $\eta$ and
the extinction coefficient $\kappa$ :
\begin{eqnarray}
\tilde k_{\alpha,\beta}^{2}=\left(k^{0}_{\alpha,\beta}\right)^{2}[(\eta_{\alpha,\beta}-i
\kappa_{\alpha,\beta})^{2}-\sin^{2} \vartheta_{\alpha,\beta}].
\end{eqnarray}
 $\eta$ and $\kappa$
are taken at room temperature as a function of the wavelength from Ref.[\onlinecite{Palik}]. 

When the density decreases at the metal surface,
the optical properties are strongly different from the above values and the electrical
conductivity $\tilde\sigma$ for dense metal plasma at the equilibrium
temperature is determined using the tabulated values given in
Ref.[\onlinecite{Ebeling}]. Experimental measurements of the electrical conductivity of dense copper and aluminum plasmas in the equilibrium temperature range of several eV and for a density range from the solid density down to 0.01 g/cm$^{3}$ \cite{DeSilva98} have been compared with theoretical models for conductivity by Ebeling \emph{et al.}\cite{Ebeling}. We used the resultant $\sigma_{DC}$ and the ionization rate under a tabulated form for given electron density and temperatures in our
calculation. A phenomenological collision frequency is deduced from these values to determine the collision frequency at equilibrium $\nu_{eq}(N_{e},T_{e}=T_{i})$. It has to be noted that an interpolation is performed between solid
and plasma phase for intermediate regime. For the liquid, gas and plasma cases,
 the dielectric function is given by the Drude model and the equation~(\ref{k0}) becomes:
\begin{eqnarray}
\tilde k^{2}_{\alpha,\beta}=\left(k_{\alpha,\beta}^{0}\right)^{2} \left(
1-\sin^{2} \vartheta_{\alpha,\beta}-\frac{i \tilde
\sigma_{\alpha,\beta}}{\omega_{\alpha,\beta} \varepsilon_{0}}
\right).
\end{eqnarray}
Here, $\varepsilon_{0}$ is the vacuum
dielectric permittivity. 

\subsection{Reflectivity and transmissivity calculations}
The reflectivity ${\cal{R}}^{p,s}$, where ($p,s$) denotes the polarisation state, is determined in a general manner, by calculating the ratio between
the reflected and the incident power. The transmissivity $\TT^{p,s}$
is evaluated by the ratio between the transmitted and the
incident power. The absorptivity is determined by the ratio between the absorbed and the
incident power. The relation
$\TT^{p,s}=1-{\cal{R}}^{p,s}-{\cal{A}}^{p,s}$ is satisfied. The angle
dependence of the incident laser absorption is an important factor
for a good optimization of the interaction. In this 
numerical experiment, the reflection factor is determined as the averaged value
integrated on the measurement time. The calculation is realised by weightening the
incident intensity with the following form :

\begin{eqnarray}\label{Rmoyen}
\overline{{\cal{R}}^{p,s}_{\alpha,\beta}}(\tau)=\frac{\displaystyle\int\nolimits_{0}^{\tau}
{\cal{R}}_{\alpha,\beta}^{p,s}(t) \bigl{|}
\EE_{\alpha,\beta}^{p,s}(t) \bigl{|}^{2}dt}
{\displaystyle\int\nolimits_{0}^{\tau} \bigl{|}
\EE_{\alpha,\beta}^{p,s}(t) \bigl{|}^{2} dt}
\end{eqnarray}

\begin{eqnarray}\label{Tmoyen}
\overline{\TT_{\alpha,\beta}^{p,s}}(\tau)=\frac{\displaystyle\int\nolimits_{0}^{\tau}
\TT_{\alpha,\beta}^{p,s}(t) \bigl{|}
\EE_{\alpha,\beta}^{p,s}(t)\bigl{|}^{2}dt}{\displaystyle\int\nolimits_{0}^{\tau}\bigl{|}
\EE_{\alpha,\beta}^{p,s}(t) \bigl{|}^{2}dt}
\end{eqnarray}
As mentioned, simulation results are obtained in performing
pump-probe numerical experiments during which the Helmholtz equations
is solved for both pump and probe pulses. In this context, the
reflectivity and the transmissivity can be associated to distinctive
wavelengths corresponding to each kind of pulse. In fact, material
optical indices evolve as a function of the corresponding wavelength in time for each pulse. In the \textsc{Esther} code,
both reflectivity and transmissivity
are calculated at each time step $\delta t$ of the simulation. This allows that, according to the specific type of experiment, these parameters are either integrated over the duration of the pulse or they give momentary values. 
In the former case, $\tau= \tau_{L}$ is
used for the integral up boundary. For experiments
providing data resolved in time below the picosecond timescale, 
$\tau= \delta t$ has been used in the previous
expressions.

\subsection{Absorption collisional aspects in stages subsequent to excitation}
Electrons interact via the bare Coulomb interaction and after time
intervals of the order of the inverse plasmon frequency
corresponding to the relevant carrier density, the interactions become
screened. Immediately after the photon absorption phase, collisions among
electrons cause loss of coherence and a hot population of
electrons is formed. Electrons and ions are initially described
by two independent populations. The electrons evolve into hot thermal
distributions characterized by an electron temperature that can
exceed significantly that of the lattice. Such a time evolution, as
the carriers equilibrate among themselves, is mainly induced by the
\emph{(e-e)} and \emph{(e-ph)} interaction. The duration of this
stage is determined by the energy relaxation time, which can be of
the order of picoseconds. The carrier dynamics affects the
absorption properties through the conductivity parameter. In a
first approximation, this quantity is completely described by the frequency-dependent
Drude contribution :
\begin{equation} \label{drude}
\tilde\sigma^{D}_{\alpha,\beta}=\frac{N_{e}e^{2}}{m_{e}^{\dag}} \frac{\nu+i\omega_{\alpha,\beta}}{\nu^{2}+\omega^{2}_{\alpha,\beta}},
\end{equation}
where the collision frequency $\nu$ is given by the sum of the
electron-electron ($\nu^{ee}$) and electron-ion ($\nu^{ei}$)
contributions. We can therefore express the collision frequency as a sum of equilibrium and nonequilibrium contributions :
\begin{eqnarray}
\nu&=&\nu_{eq}+\nu_{neq} \nonumber\\
&=&\nu_{eq}(N_{e},T_{i})+\nu_{neq}(N_{e},T_{e}) 
-\nu_{neq}(N_{e},T_{e}=T_{i})
\label{nu}
\end{eqnarray}
Here, $\nu_{eq}(N_{e},T_{i})$ and $\nu_{neq}(N_{e},T_{e}=T_{i})$ are \emph{(e-i)} collision frequencies at equilibrium, calculated in two different ways. The first term is determined using the tabulated values of conductivity at equilibrium \cite{Ebeling}. These tabulated values were shown to match well experimental situations  \cite{DeSilva98}. The last term corresponds to the equilibrium contribution to the collision frequency calculated by our nonequilibrium model. To ensure a convergence towards tabulated equilibrium values, we substract the equilibrium contribution from the nonequilibrium one. In this manner, the contribution of electrons out of equilibrium is treated as a correction of the known values of the equilibrium conductivity. Finally, the second term, $\nu_{neq}(N_{e},T_{e})$, is the nonequilibrium collision frequency for electrons which account for \emph{(e-e)} (via umklapp processes) and \emph{(e-i)} interactions for an electronic temperature larger than the ionic one. This later contribution is calculated independently from the former one as it will be seen in the next paragraph. 

The electron-electron part is a fundamental issue for the description of 
low-temperature or nonequilibrium transport phenomena. It is a difficult problem to deal correctly with electron-electron
scattering in calculations of transport coefficients. One of the possible way is the Boltzmann transport equation. However, the
Boltzmann transport equation cannot usually be solved exactly
and some approximations have to be introduced, such
as a relaxation-time approximation. No tractable model exists to describe
the conductivity on a large range of temperatures and densities and
we propose an interpolation between the different regimes with established characteristics, the solid and the plasma states.
 To take this into account, we have used an approach based on an interpolation between current
solid and plasma model following approach in Ref.[\onlinecite{Fisher01}]. 
The relaxation rate ($1/\nu$)
is calculated with the assumption of a screened Coulomb interaction among the electrons.
The \emph{(e-e)} relaxation rate rises with increasing density until it reaches a maximum after
which increasing screening causes the rate to fall. The evolution of the optical
properties during an ultrashort laser pulse remains an
open question. A complete description of the effect of the
\emph{(e-e)} collisions would require the complete knowledge of
both the metal band structure and dense-plasma atomic structure,
which makes them analytically complex and unsuitable for the
modelling purposes considered here. In reality, it is not quite obvious how electron-electron collisions can induce a conductivity effect in the nonequilibrium solid. For a free electron gas, the parabolic energy-momentum relation for which velocity and momentum are proportional results in neglegting \emph{(e-e)} interactions contribution to absorption. The classical approach implies that, since total current and total momemtum are proportional, the current is conserved during this type of collisions which preserves momentum. As a consequence, the conductivity evolution is supposed to be independent of \emph{(e-e)} interactions. However, in periodic systems, as the solid sample in the first moments of interaction, the total momentum involving \emph{(e-e)} interactions may be conserved via the involvement of a reciprocal lattice vector. This type of process will allow in consequence a variation of the electronic momentum and induces a change in conductivity \cite{Abrikosov}. A term $\nu^{ee}_{neq}$ is then used to correct the conductivity during the nonequilibrium stage. To justify our approach, we expect that umklapp processes can occur at a rate given by a factor $\Delta$ with respect of the normal process ($\Delta<1$) \cite{Lawrence72,Parkins81}. The values of this term ($0.4$ and $0.35$ for aluminum and gold, respectively) have been taken from Refs.[\onlinecite{Macdonald80,Kaveh84,Groeneveld95}], being derived by low-temperature investigations. These values will be considered constant in this study, although a dependence on the electronic field and collision rates is foreseeable. In taking into account this process, the slowing down of the electron flow can be expected during \emph{(e-e)} interaction and results in a change in the absorption coefficient when the lattice remains cold, during the first moments of laser irradiation. 

At near-solid density and moderate electronic temperature up to few
eV, the Fermi gas undergoes collisions and the solid-model is
given by~\cite{Voisin04} :
\begin{equation}
\nu_{neq}^{ee}=\frac{\CC}{\cal{N}}
  \int_{0}^{\infty} d{\cal{E}} ({\cal{E}}-{\cal{E}}_{F})^{2} f( {\cal{E}},T_{e} )[1-f({\cal{E}},T_{e})]
\end{equation}
where $f( {\cal{E}},T_{e} )$ is the Fermi-Dirac statistic
distribution depending on $T_{e}$ and $N_{e}$ through the chemical
potential.
\begin{eqnarray} \label{Voisin}
\begin{cases} {\cal{N}} = -k_{B}T_{e}f(0,T_{e})\\
\CC=\Delta \displaystyle\frac{\DD}{{\cal{E}}_{TF}^{3/2}
{\cal{E}}_{F}^{1/2}  } \left[\frac{2\sqrt{{\cal{E}}_{TF}
{\cal{E}}_{F} }}{4 {\cal{E}}_{TF} + {\cal{E}}_{F}
}+\arctan{\sqrt{\frac{4{\cal{E}}_{F}}{{\cal{E}}_{TF}}}} \right] \\
{\cal{E}}_{TF}=\displaystyle\frac{\hbar^{2}q_{TF}^{2}}{2m_{e}^{\dag}} \
\text{where} \ q_{TF}=\displaystyle\frac{\zeta e}{\pi \hbar}
\sqrt{\frac{m_{e}^{\dag}}{\epsilon_{0}\epsilon_{d}}}
(3\pi^{2}N_{e})^{1/6}
\end{cases}
\end{eqnarray}
where $\DD=m_{e}^{\dag} e^{4}/64\hbar^{3}\pi^{3}\epsilon_{0}^{2} \epsilon_{d}^{2}$, 
$m_{e}^{\dag}$ is the effective electron mass \cite{Palik}, $\Delta$ is the rate of umklapp processes, 
 and $\epsilon_{d}$ represents the contribution
of band $d$ electrons to the static dielectric function in the
case of noble metals. In the same way as described in Ref.[\onlinecite{Voisin04}], we have used the screening wave-vector $q_{TF}$ containing the adjustable parameter of screening $\zeta\leq1$
 which was set by fitting the experimental results of these authors. The solid model is applicable up to the Fermi
temperature. We assume that the absorption
is mainly driven by the free electrons and this model allows us to calculate a
straightforward dependence in $T_{e}$. This approximation provides a schematic view of the
absorption which can be included in the code. For
electrons at temperatures around the Fermi value, the phase space available
for scattering increases with the electronic temperature due to the Pauli exclusion. Note that the increase in the \emph{(e-i)} collision frequency due to the $T_{e}$ augmentation is supposed to be negligible with respect to the \emph{(e-e)} contribution.

For larger electronic temperatures, we use the plasma collision frequency
given by the Spitzer formula:
\begin{equation}
\nu_{neq}^{ei}=\frac{{\cal{E}}_{F}}{\hbar}\left(\frac{k_{B}T_{e}}{{\cal{E}}_{F}}\right)^{-3/2}
\end{equation}

The approximations inherent to the plasma model are more acceptable at
low density compared to using the solid model at high temperature
\cite{More88}. To reduce the interpolation range in the unknown
region, we suppose that the Spitzer assumptions become suitable
for $T_{e}>2T_{F}$. These assumptions may be criticized because of the high density close to the solid one, but the choice of this direct temperature dependence is largely used to represent the global behavior.
The nonequilibrium frequency over a large range of temperature is
then defined by:
\begin{eqnarray}
\begin{cases}
\nu_{neq}=\nu_{neq}^{ee}(N_{e},T_{e}) \ \text{for} \  T_{e}\leq T_{F} \\
\nu_{neq}=\nu_{neq}^{int}(N_{e},T_{e}) \ \text{for} \  T_{F}<T_{e}<2T_{F} \\
\nu_{neq}=\nu_{neq}^{ei}(N_{e},T_{e}) \ \text{for} \ T_{e}\geq
2T_{F}
\end{cases}
\end{eqnarray} where $\nu_{neq}^{int}=a T_{e}^{3}+b T_{e}^{2}+c
T_{e}+d$ is a cubic interpolation which offers a realistic
continuity between the two models. The interpolated data
points are created so that they have a slope equal to the slope of
the start or end segments. Note that each term ($a$,$b$,$c$,$d$)
is calculated and tabulated as a function of $N_{e}$.

\begin{figure}[H]
\begin{center}
\includegraphics[width=8.5cm]{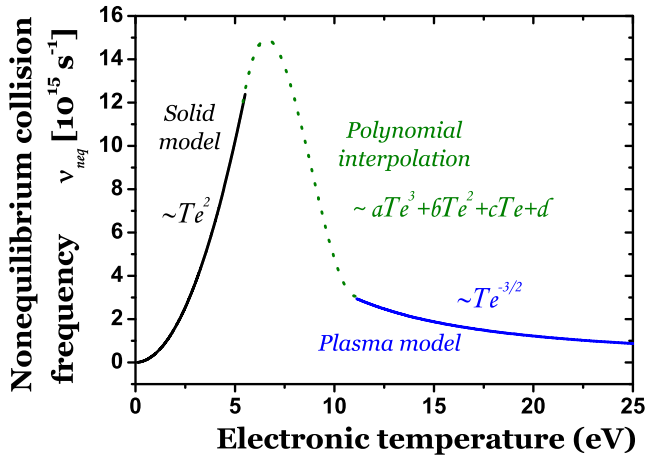}
\caption{Nonequilibrium collision frequency interpolated in a wide range of electron temperatures in Au. Solid and plasma models
(solid lines) are connected by a polynomial interpolation (dashed
line) which allow extension in the unknown
regime.}\label{interpol}
\end{center}
\end{figure}

As a conclusion, direct \emph{(e-e)} contribution to the optical absorption are taken into account via umklapp processes which are supposed to occur in the solid phase via a reciprocal lattice vector. This allows, in turn, a variation of the electronic momentum, and, subsequently, a contribution to the optical conductivity. Concerning \emph{(e-i)} contribution, the indirect screening of the ion potential by electronic influences is accounted for by the popular Spitzer dependence of the nonequilibrium collision frequency. The equilibrium conductivity values are derived from theoretical considerations which match experimental conductivity results. We suppose that the electric field influence is taken into account indirectly by the effect of the electronic energy on the collision rate. In this way, the nonequilibrium ($T_{e} \ll T_{i}$) affects the Joule heating in the condensed phase in considering the contribution of \emph{(e-e)} collisions. Although interpolation between solid and plasma regimes is questionnable, we suppose that \emph{(e-e)} effects prevail during the laser-solid interaction, yielding a dramatically modified scaling $\nu_{neq}\propto T_{e}^{2}$\cite{Martinolli06}. A strong influence of this law is expected on the laser absorption, which will be discussed in the next sections.

\subsection{Interband contribution}
The absorption  model was implemented for two metallic materials, gold and aluminum. For gold, no specific model of nonequilibrium interband contribution was implemented. We assume that the equilibrium contribution of the collision frequency already contains interband aspects. Moreover, it is supposed that the contribution of $d$ bands in the case of noble metals was taken into account in equation (\ref{Voisin}).

For aluminum, the dielectric constant of the metal takes into account two
mechanisms of laser energy deposition: intraband
absorption in the $sp$ band and interband absorption (parallel
band). Intraband absorption is described through the Drude model, while the
Ashcroft-Sturm model is used for interband contribution~\cite{Ashcroft71}. The influence of
interband transition in aluminum conductivity is then considered as $\tilde\sigma=\tilde\sigma^{D}+\tilde\sigma^{IB}$.
Fisher \emph{et al.} have reported theoretical and experimental investigations on aluminum absorption for femtosecond laser irradiation in
showing the influence of the interband and intraband (Drude) contributions at 400 and 800~nm~\cite{Fisher01}. These results were verified in conducting a comparable model of interband effects on the conductivity and similar results were obtained by our numerical code.

\section{Experimental validation of theoretical model}
In order to validate our simulations with experimental results,
this section is dedicated to a comparison between selected literature data
and numerical calculations based on the present model of ultrafast
laser absorption. The experimental part focuses on two demonstrations: two kinds of experiments were reproduced to
investigate the optical properties of high density plasmas.
Studies on absorption dependence in angle of incidence and intensity
were performed. Moderate and high intensities were
considered for UV ultrashort laser pulses. The results provide optical
properties in both solid and high density plasma range during the
laser pulse.

\begin{figure}[htbp]
\begin{center}
\includegraphics[width=8.5cm]{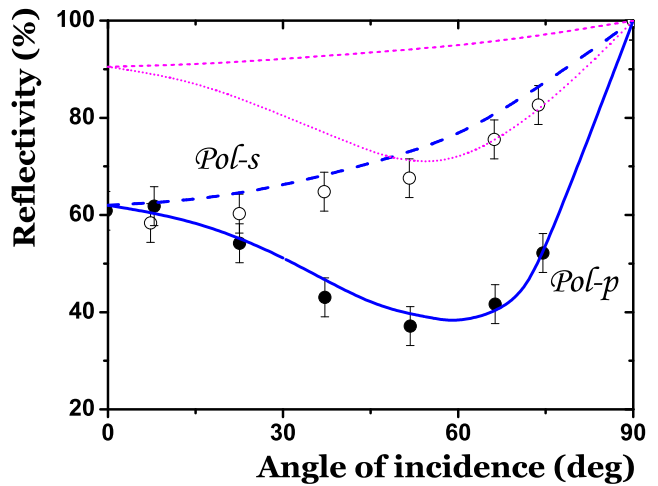}
\caption{Simulated reflectivity as a function of angle of
incidence for a 250~fs, $248$~nm laser pulse for $s$-polarised
(dashed line) and $p$-polarised (solid line) irradiation of an
aluminum target at $10^{14}$ W/cm$^{2}$ intensity. Minimal
reflectivity is obtained for an angle of $(59\pm3)^{\circ}$.
Simulations performed with $\nu_{neq}=0$ are superposed for $s$ (short-dashed line) and $p$ (dotted line) polarization.
Experimental data points (open ($s$) and closed ($p$) circles with
error bars) are taken from Fedosejevs \emph{et
al.}\cite{Fedosejevs90}.\label{angle}}
\end{center}
\end{figure}

 Fedosejevs \emph{et al}. \cite{Fedosejevs90} have measured the
averaged reflectivity corresponding of a KrF laser pulse of 250 fs at $248$
nm focused on aluminum samples as a function of the polarization
and the angle of incidence for two intensities: $10^{14}$
W/cm$^{2}$ and $2.5\times10^{15}$ W/cm$^{2}$. At the higher
intensity, the pre-pulse is supposed to be sufficiently intense to
produce a pre-plasma at the metal surface. In fact, the absorption
of the main pulse is strongly modified by the pre-plasma
occurence. In this condition we restrain our study to the lower
intensity experiment, which corresponds to a laser fluence of 25
J/cm$^{2}$. The experimental measurements were performed to
demonstrate the characteristic angular dependence of
electromagnetic absorption expected for the extremely steep
density gradients of plasmas. Our calculations can be
compared with the experimental data since we solve the wave
equation for non-stationary density and temperature profiles. If
the absorptivity decreases with the angle of incidence for the
$s$-polarization, a maximum absorptivity around $62\%$ for the
$p$-polarization was observed for an angle of $(54 \pm
3)^{\circ}$. This corresponds to the maximum energy deposited in the material. In the experiment, the incident beam has a
non-desired mixed polarization and the intensity corresponding to
a $p$-polarization is in fact composed of 93~\% $p$-polarization
and 7~\% $s$-polarization. The inverse ratio corresponds to
$s$-polarization. This polarization state was entered in our
simulations and the calculated reflectivity is given as a function
of the parameter of polarization $x_{p}$ which has been fixed to
be equal at $0.86$ in the $p$-polarization case and $-0.86$ in the
$s$-polarization case:

\begin{equation}\label{polar}
\overline{{\cal{R}}_{\alpha}}=\frac{1}{2}
\left[(1+x_{p})\overline{{\cal{R}}_{\alpha}^{p}}+(1-x_{p})\overline{{\cal{R}}_{\alpha}^{s}}\right]
\end{equation}
Fig.~(\ref{angle}) shows the reflectivity as a function of the
angle of incidence for both polarizations. The simulated values
were calculated with the expressions (\ref{Rmoyen}) and (\ref{polar})
with $\tau_{L}=750$ fs corresponding to the total
pulse duration. The simulations were performed with the
expression~(\ref{nu}) for dominantly $p$-polarization (solid curve) and
$s$-polarization (dashed curve). To show the influence of the
nonequilibrium absorption term and especially the \emph{(e-e)}
collision effects, other simulation results are represented by
setting the nonequilibrium contribution to $\nu$ equal to zero. 
The expression~(\ref{nu})
is reduced to the first equilibrium term which only depends on $N_{e}$
and $T_{i}$ ($\nu_{neq}=0$).
These simulations have been done for $p$-polarization
(short-dotted curve) and $s$-polarization (short-dashed curve).
The reflectivity values in this later case are higher than the
experimental ones and the simulations that include $\nu_{neq}$ are clearly more relevant. On the
 subpicosecond timescale,  the ionic temperature does not have enough
time to increase because the energy transfer rate between
electrons and ions has a characteristic time longer than
$\tau_{L}$. For the case $\nu_{neq}(N_{e},T_{e}) =0$, the lattice remains cold during the laser pulse and
the conductivity, depending on the electron-ion collision
frequency, is close to the cold solid one. The $p$-polarization shows a minimum reflectivity at closed to $59^{\circ}$.
This minimum which appears only for the $p$-polarization case, can be explained by 
the fact that for this polarization, the electromagnetic wave has a component colinear to the
plasma density gradient. 

The minimal reflectivity occurs for a calculated angle of approximately $59^{\circ}$ which corresponds roughly to the experimental one. Eidmann~\emph{et al}. have already compared these experimental results with their hydrodynamic simulations which solve Helmholtz equation in a similar framework~\cite{Eidmann00}. Nevertheless, their model does not consider contributions of \emph{(e-e)} collision in the solid case but the \emph{(e-ph)} contribution was calculated with a different specific plasma model \cite{Eidmann00}. In this way, the collision frequency increased quickly during the laser pulse and they have found an angle corresponding to the minimal absorption around 70$^{\circ}$ in their calculations. They have imputed this discrepancy to the fact that they had not considered the depolarization in the calculation of the electromagnetic field. We have observed the same situation, where the angle of maximal absorption was overestimated with a pure $p$-polarized pulse.
In our calculation, this overestimation was corrected, by including a weak proportion of $s$-polarized wave in
the total electromagnetic field. These simulations reveal a strong difference in the nature of the absorption
of the $s$ and $p$ components of the heated matter absorption. When a density gradient is created at the metal surface,
simplest models used to calculate the electromagnetic fields at two interfaces or in the skin depth are no longer valid and
it becomes necessary to solve the Helmholtz equation to calculate the deposited energy. Fedosejevs \emph{et al.}
have already indicated the importance to take into account a correct density gradient to reproduce the experimental data. One dimensionnal
modeling appears to be sufficient to take into account the effects due to incident angle and polarization. Concerning the
magnitude of the reflectivity, we have pointed out the crucial role of the correct conductivity by way of a total collision frequency
including the \emph{(e-e)} collision rate. Moreover, the fact that the angle of minimal absorption was similar to the
experiment, show that hydrodynamic of the system is well reproduced by our simulations
on the subpicosecond timescale. 

In order to test our model on a second type of study, we have investigated the laser intensity dependence of these two polarisations while keeping constant the angle of incidence. The dependence of the reflectivity
as a function of the intensity was experimentally studied by Milchberg~\emph{et al}.
to determine the evolution of the aluminum resistivity during the solid-plasma transition when the change in density is weak~\cite{Milchberg88}.
These measurements were interpreted to provide an electronic temperature dependence for the resistivity. This dependence was studied for an aluminum target with a density close to the solid one for an intensity evolving for four orders of magnitude, from $5 \times 10^{11}$ to $10^{14}$ W/cm$^{2}$. The reflectivity was determined by measuring the energy after reflexion of the laser pulse
which hits the target at an incident angle of $45^{\circ}$. The target were prepared by depositing a 400~\AA~aluminum film on a glass substrates. The authors stated that for thicker film, the reflectivity remains constant.

\begin{figure}[htbp]
\begin{center}
\includegraphics[width=8.5cm]{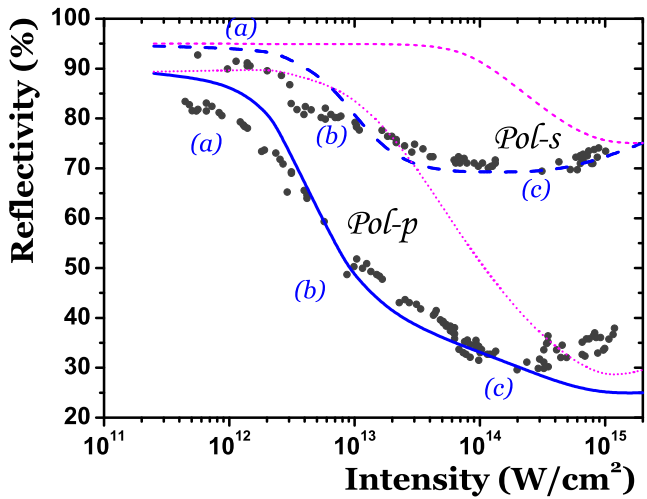}
\caption{Simulated reflectivity as a function of intensity in an
aluminum sample for $s$ (dashed line) and $p$ (solid line) polarization.
Simulations performed with $\nu_{neq}=0$ are superposed for $s$ (short-dashed line) and $p$ (dotted line) polarization.
  Experimental data points (black dots) are taken from Milchberg \emph{et
al.}\cite{Milchberg88}.}\label{Milchberg}
\end{center}
\end{figure}

Fig.~(\ref{Milchberg}) shows the experimental reflectivity
associated to the 400 fs (FWHM) $s$- and $p$-polarized incident pulses for
fluences from 20 mJ/cm$^{2}$ to 2 kJ/cm$^{2}$ at
$\lambda_{\alpha}=308$ nm. A decrease of reflectivity with intensity is  observed. The challenge is to relate this decrease to an excitation model. The figure also shows the calculated result in two situations ($\nu_{neq}=0$, $\nu_{neq}\neq0$). Our simulations are
in good agreement with these experimental measurements when the \emph{(e-e)} collisions are considered. Numerical
calculations provide reflectivity evolutions based on the
expression~(\ref{Rmoyen}) where the upper limit of the integration
time corresponds to the total pulse duration $\tau_{L}=1.2$ ps. A drop of reflectivity for an intensity range of $10^{12}-10^{15}$ W/cm$^{2}$ is observed. In the case $\nu_{neq}=0$,
the dependencies in $T_{i}$ and $N_{e}$ on the \emph{(e-i)} collisional processes are not sufficient to reproduce
the experimental data and the agreement is poor. The drop in reflectivity occurs for higher intensities for both polarizations  when the
increase in ionic temperature starts to be important and to drive the absorption mechanism. This again shows the role of considered \emph{(e-e)} collision in influencing the transient optical properties.

Three regimes are clearly distinguishable when observing reflectivity as a function of the
laser intensity. The reflectivity can be written as a function of
the refractive index $\tilde n$ as ${\cal{R}}=|(1-\tilde n)/(1+ \tilde n)|^{2}$, where,
according to the Drude expression:
\begin{equation}
\tilde n^{2}(\omega_{\alpha}/\nu)=1
+\frac{\omega_{p}^{2}}{\omega_{\alpha}^{2}} \frac{i
\omega_{\alpha} /\nu}{1-i \omega_{\alpha}/\nu},
\end{equation}
where $\omega_{p}=(N_{e}e^{2}/m_{e}^{\dag}\varepsilon_{0})$ is the plasma frequency. The above equation used in the case of normal incidence of laser
radiation, and the reflectivity at other incident angles for $p$- and
$s$-polarized radiation can be calculated based on the
Fresnel formulas. These considerations do not bring new informations for a quantitative discussion but allows a qualitative one.
If $1\leq \omega_{\alpha}/\nu\leq  \omega_{p}/\nu$ with $\omega_{\alpha}=6.12\times
10^{15}$s$^{-1}$, then $\tilde n^{2}\simeq 1+i\omega_{p}^{2}/\omega_{\alpha} \nu$ and $\overline{{\cal{R_{\alpha}}}}$ decreases with $\nu$,  corresponding to the first regimes. The first one, indicated by the label $(a)$ is observable at intensities lower than few $10^{12}$ W/cm$^{2}$. This regime,
caracterized by a reflectivity which decreases slowly, corresponding
to an electronic temperature which is not sufficiently high to
induce high collision frequency ($\nu\ll\omega_{\alpha}$). Moreover, the energy exchange between electrons and ions is low and the matter remains in a solid state during the pulse duration. The  $\overline{{\cal{R_{\alpha}}}}$ decrease is related to the rise of \emph{(e-i)} collision frequency, which depends almost linearly on the ionic temperature. Moreover, the decrease in density is weak in this regime and the reflectivity does not depend on the electronic density through $\nu$ and $\omega_{p}$. The second regime, marked by the label $(b)$ on the Fig.(\ref{Milchberg}), is reached for intensities between $10^{12}$ and $10^{14}$ W/cm$^{2}$. In this case, $\nu\simeq \omega_{\alpha}$ and the drop of $\overline{{\cal{R_{\alpha}}}}$ can be related to a strong $T_{e}$ increase which leads to a frequency collision augmentation. In the assumption of the
solid-model, $\overline{{\cal{R_{\alpha}}}}$ decreases as $\nu_{neq} \propto T_{e}^{2}$, which corresponds to the regime put forward by Milchberg
\emph{et al.} in Ref.[\onlinecite{Milchberg88}]. Note that the decrease in density starts to play a significant role in this regime as well and can induce a reduction of the reflectivity. Finally, for the highest intensities, the reflectivity remains constant or increases.
This regime is more difficult to reproduce with simulations because, beyond $10^{14}$ W/cm$^{2}$ the experimental pre-pulse
is sufficient to produce vapor or plasma before the absorption of the main pulse. In this case marked by the label (c), the electronic critical density reached in front of the aluminum sample yields a strong
modification of the absorption properties.
For the $p$-polarization case, $1\ll \omega_{p}/\nu \leq  \omega_{\alpha}/\nu$ and $\overline{{\cal{R_{\alpha}}}} \simeq (\omega_{p}/2\omega_{\alpha})^2$ becomes low, depending on the electronic density.

The good agreement between the experimental and theoretical reflection factor
shows the possibilities of using the presented model to reproduce and analyse
the optical properties during the plasma formation induced by a laser pulse. This was achieved by properly estimating the collision frequency and other optical and thermodynamical parameters in nonequilibrium conditions. These results indicate that the numerical code is able to simulate the absorption of the incident electromagnetic wave in electron-ion nonequilibrium
regimes. This provides opportunities to caracterize the response of a material for timescales lower than
the electron-ion relaxation time. We have studied reflectivities induced on a thick metal sample,
where the heated matter at the surface was composed by a temporal sequence of several layers of different
thermodynamic states. The next section is dedicated to specific experiments which allow to follow a single
thermodynamic state on ultrashort timescales ensuring as well temporal resolution. Our simulations are suitable to provide
precious informations on the collisional processes and state of the system during its relaxation.

\section{Application to pump-probe diagnostic}
Optical reflectivity and transmission measurements provide valuable insights into studying ultrafast phenomena in condensed matter and plasmas physics \cite{Ng98}. In particular, they enable observations of phase transitions
 and provide information about the transport properties of
dense plasmas. If the resolution is short enough, pump-probe
experiments are relevant to characterize the properties of the
material during relaxation processes. The use of ultrashort laser
pulse provide a sub-picosecond resolution of the
evolution after an ultrafast laser event, the pump excitation. The
subsequent dynamics is then governed by the interactions among the
elementary excitations and is monitored by using a second optical
pulse, the probe. Reflectivity measurements strongly depend on
plasma density and collision frequency. In these conditions, numerical
investigations can provide the correlation between matter
properties and relaxation process. In this section, we compare
experimental results and simulations performed in similar
conditions which permit us to split up the sequence of
electron-electron, electron-phonon and hydrodynamics relaxations.

\subsection{Strong pulse and unperturbing probe}
Our assumptions are firstly tested against the data points corresponding to pump-probe reflectivity and transmission experiment
performed by Widmann~\emph{et al}. in Ref.~[\onlinecite{Widmann04}]. In this case, the target is a gold foil with a
thickness of about 28 nm which has been irradiated with a $\lambda_{\alpha}=400$~nm,
150~fs (FWHM) laser pulse, at normal incidence. After
focusing, the intensity of the pump pulse is about
$10^{13}$~W/cm$^{2}$ on the metal surface. The reflectivity and
the transmission of the initial solid and the produced plasma are
analyzed with a \emph{s-}polarized probe pulse at $\lambda_{\beta}=800$~nm with a
similar duration, an incident angle of 45$^{\circ}$, and a time delay
varied from 0 to 15 ps after the pump pulse. In the experiment,
reflected and transmitted energies were measured with photodiodes
and compared to the incident energy to deduce an instantaneous
reflectivity and transmission of the matter. In a similar way as in the previous section,
simulated reflectivity $\overline{{\cal{R}}^{s}_{\beta}}(\delta
t)$ and transmission $\overline{\TT^{s}_{\beta}}(\delta t)$ were calculated with the expressions~(\ref{Rmoyen})
and~(\ref{Tmoyen}) where $\tau$ was taken to be equal to a
variable timestep $\delta t \leqslant 0.1\ fs$.

\begin{figure}[htbp]
\begin{center}
\includegraphics[width=8.5cm]{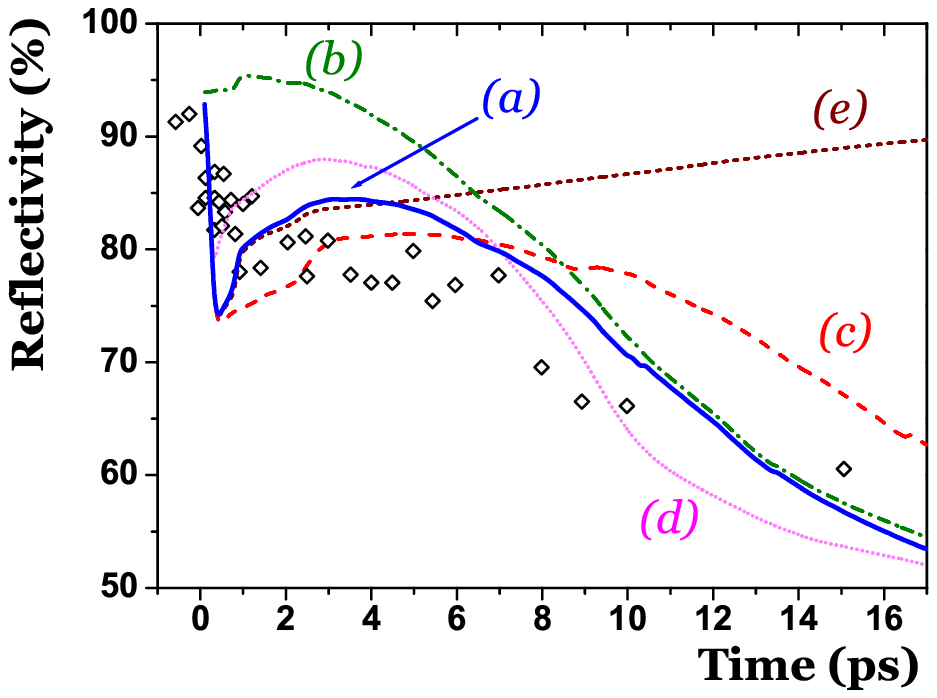}
\includegraphics[width=8.5cm]{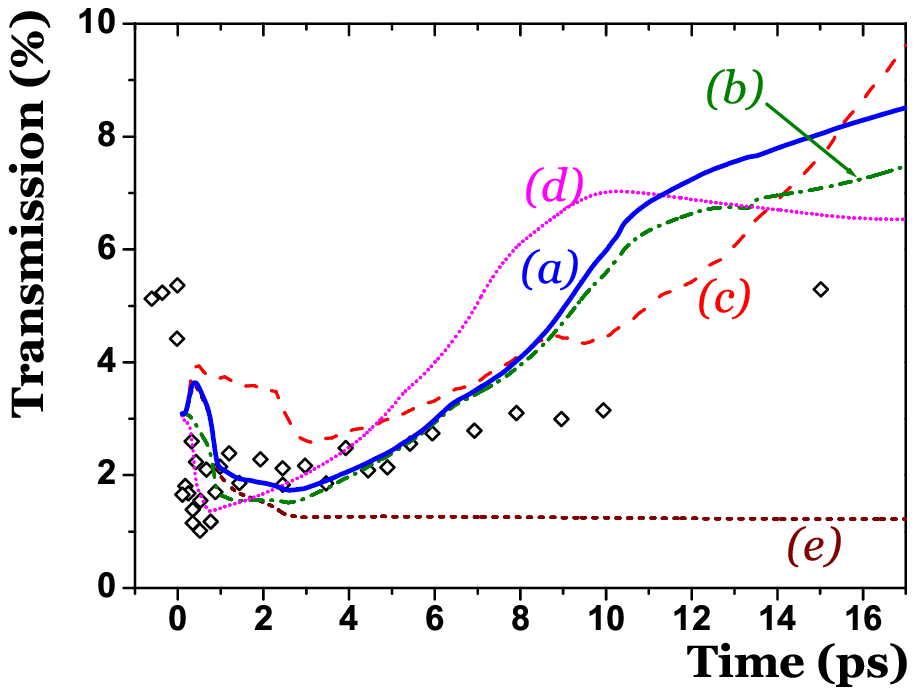}
\caption{Numerical evolution (curves) of the reflectivity and the
transmission of a 800~nm pulse probing a 28~nm gold foil
thickness, for different simulation configurations. The solid line
(a) corresponds to the best-fitted complete simulation. The
dash-dotted line (b) is a simulation without nonequilibrium
contribution in the absorption. The dashed line (c) and the
short-dotted line (d) correspond to simulations performed with a
lower ($\gamma=\gamma_{0}/4$) and a higher
($\gamma=\gamma_{0}\times 2$) \emph{e-ph} coupling term,
respectively. The hydrodynamics processes have been switched off in
the simulation shown by the short-dashed line (e). Experimental
data points (black diamonds) are taken from Widmann \emph{et
al} \cite{Widmann04}. \label{Ng}}
\end{center}
\end{figure}

At the beginning of the irradiation and at the probe frequency,
solid gold shows a reflectivity of $93\%$. Following the
irradiation, three regimes are clearly distinct on the
Fig.~(\ref{Ng}). The first one, shortly after the initial moment,
is characterized by a swift decrease in both reflectivity and
transmission during the first picosecond. The second one is marked
by a plateau between 1-7 ps and corresponds to a regime where
optical properties seems to be frozen because reflectivity and
transmission are almost constant. Finally, a change appears after 7
picoseconds and a last regime starts, characterized by a decrease
in reflectivity and an increase in transmission. To explain these
behaviors, we have performed five simulations with different
configurations to identify the effect of the supposed main
processes. 

The first simulation labelled by $(a)$ on the
Fig.~(\ref{Ng}) is a complete simulation including all the
processes described in the section II and is used as the reference
case to interpret the other calculations. The dash-dotted line
$(b)$ is a simulation without nonequilibrium contribution in the
collision frequency. In this case, $\nu_{neq}$ contribution has been
skipped from the equation~(\ref{nu}) and no \emph{(e-e)} collision
mechanism is taken into account in the bremsstrahlung absorption
calculation. The expected difference has to take place during the
nonequilibrium process, before electron-phonon thermalization.
The calculated reflectivity is strongly different in the case
$(b)$ since the conductivity parameter only depends on the density and on the
ionic temperature which evolves slowly compared to the electronic
temperature variation. In contrast, curve $(a)$ shows a
drop in the reflectivity and reaches a minimum value at the end of the laser pump pulse, when the
electronic temperature is maximal. At this time, $\Re_{e}[\tilde\sigma_{\beta}]$ and the subsequent quantity of absorbed energy are maximal. 
The transmission curve displays
a less strong difference for these two cases because $\overline{\cal{R_{\beta}}}$ diminution and absorption increase occur simultaneously. For the case $(a)$, as electrons transfer their energy to the ions, $\nu$ decreases and $\overline{\cal{R_{\beta}}}$ rises and reaches a maximal value at 4 ps. From this study, we can infer that the
observed absorption in ultrafast timescales is mainly due to
\emph{(e-e)} collision. Thus, during the electromagnetic
excitation, the Coulomb interaction between electrons drives the Joule
heating process.

The curves $(c)$ and $(d)$ include nonequilibrium
absorption mechanism but differ from $(a)$ in \emph{(e-ph)} energy
transfer rate. In fact, the case $(c)$ corresponds to a case where
the usual \emph{(e-ph)} coupling term $\gamma_{0}=4\times 10^{16}$ W
K$^{-1}$ m$^{-3}$ has been divided by a factor 4 to slow down the
ionic temperature increase and to increase the nonequilibrium
lifetime. Compared to the experimental data and the $(a)$
reference case, the break-up with the third stage at 7 ps is
smoothened and the region where the reflectivity remains constant is
more pronounced. In contrast, for the $(d)$ case, the
\emph{(e-ph)} coupling term $\gamma_{0}$ was multiplied five times to 
accelerate the ionic temperature increase and to
decrease the nonequilibrium lifetime. In this case, the length of
the plateau (which is visible in both reflectivity and transmission)
decreases and the equilibration time seems to be reduced. We
deduce that the plateau can be attributed to the temperature
equilibration mechanism between electrons and ions and the
duration of this stage corresponds to the \emph{(e-ph)} relaxation
time. After this, the hydrodynamics commences.

Finally, the fifth round $(e)$ was performed by
keeping $\gamma=\gamma_{0}$ as in \emph{(a)} but the hydrodynamics
processes have been stopped. To do this, the pressure $p$ and the
density $\rho$ equal the standard pressure $p_{0}=10^{-5}$ Pa and
the solid density $\rho_{0}=19.3$ g~cm$^{-3}$, respectively, in the entire gold film.
 In this non-physical case, no pressure gradient
can cause motion, volume increase, surface expansion, or shock
processes in the solid. The behavior of the optical properties is
similar for $(e)$ and $(a)$ during the 3-4 first picoseconds and, after this time, the reflectivity and the transmission split and
follow very different evolutions. In the isochoric case,
the reflectivity increases continuously as the energy is
transferred from the electrons to the ions. In fact, the total collision frequency $\nu$
 decreases because the augmentation in \emph{(e-i)} collision is not compensated by the 
 \emph{(e-e)} collision decrease. 
The transmission does not evolve anymore after the end of the
nonequilibrium process since the change in reflectivity is
compensated by a weaker absorption of the probe. In the
hydrodynamic case $(a)$, the reflectivity decreases because of
the density decrease which is related to the plasma formation at the surface. In fact, according to the Spitzer
formula and the Drude model for $\nu<\omega_{\beta}$, $\overline{\cal{R}_{\beta}}$
decreases with the ionic temperature increase due to the increase in
\emph{(e-i)} collision frequency and $\overline{\cal{R}_{\beta}}$ decreases when
the electron density drops. The change in $\tilde\sigma_{\beta}$ yields a greater skin depth for the probe, leading to
increases in optical path length. As the plasma expands, the
temperature and the density falls and the free electrons
recombine.

\begin{figure}[htbp]
\begin{center}
\includegraphics[width=8.5cm]{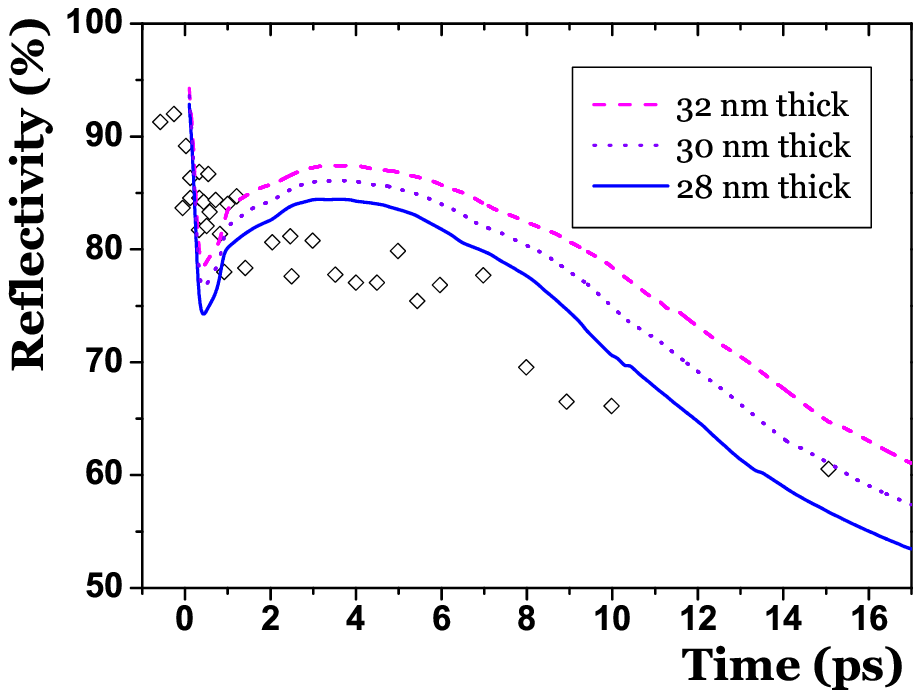}
\includegraphics[width=8.5cm]{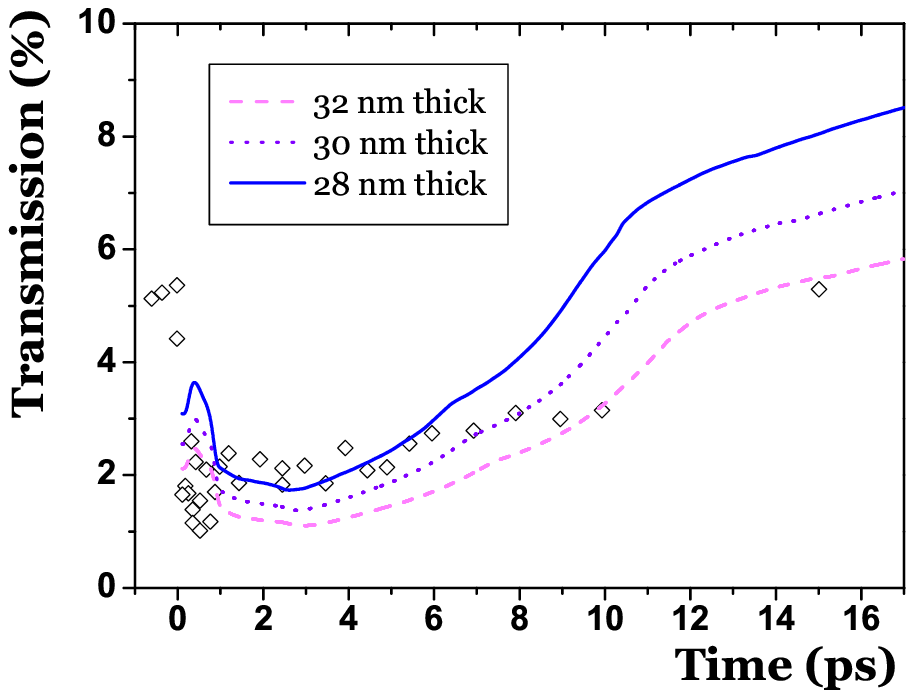}
\caption{Time evolution of the reflectivity and
the transmission of a 800~nm probe pulse for different thicknesses
of a gold sample pre-excited by a 400~nm laser pulse. Both behaviors exhibit energy relaxation stage and solid to plasma transformation induced by a
400~nm, $0.5$ Jcm$^{-2}$ pump pulse. Experimental data points
(black diamonds) are taken from Widmann \emph{et
al} \cite{Widmann04}. Generation of nonequilibrium, relaxation to equilibrium and solid to plasma transition are visible. \label{Ng2}}
\end{center}
\end{figure}

The experimental measurements were performed on $30\pm2$~nm gold sample. The magnitude of absorption strongly depends on the material thickness. 
To estimate the influence on our simulations, Fig.~\ref{Ng2} shows a dependence of the transient optical properties in the material thickness.
 In our simulation, an increase of energy of about
$4.8\times 10^{6}$ J/Kg notified by Ref.[\onlinecite{Widmann04}] corresponds to an
input fluence of $0.5$ Jcm$^{-2}$ for the pump pulse incident on a sample of 28 nm thickness. The behavior of reflectivity and transmission were investigated
for three thicknesses: 28, 30 and 32 nm at the same fluence. These three simulations show that the general evolution of the optical properties are identical but differs
in magnitude. The reflectivity increases whereas the transmission decreases with the film thickness. This simple result can give an explanation
to the discrepancy between the experimental and simulated values in providing a kind of error bar to the numerical experiment data.

This numerical study emphasizes the role of the \emph{(e-e)} umklapp collision mechanism on the absorption of the laser pulse. The \emph{(e-e)} interaction influence the material optical reponse during a short time after the pump pulse. The following \emph{(e-i)} equilibration
and the hydrodynamic stages have been identified as well. Due to the number of approximations in the optical model, the simulation does not perfectly match the experimental data. Nevertheless, the different phases of the
 matter evolution and the subsequent changes in the behavior of the optical properties are quite well reproduced. This numerical study devoted to reflect the experimental conditions involving laser-metal irradiation for both pump and probe pulses is an efficient tool to discriminate collisionnal processes. A direct interest of a low energy probe pulse is the possibility of carrying out tests of optical absorption of the material. This study could be extended to perform temporal optimisation of laser absorption in determining the time when the matter state is the more prepared and able to respond to the electromagnetic excitation.

\subsection{Experiments under near-threshold laser excitation on solid samples}

The final discussion relates to a class of pump-probe experiments which focus on particle emssion. We propose a direct application of a double-pulse sequence in studying the consequence of transient optical properties on energy coupling efficiency as a function of delay.  Experiments based on ion signal emission as a function of the time delay were performed on several metals~\cite{Schmidt00}. The experiments map in time the energy deposition as a function of the delay between identical sub-threshold pulses, the total energy being above the threshold. Here, the threshold is defined as ion emission threshold. This way a quantitative picture of the transient behavior of the energy coupled into the sample emerges. It is supposed that the energy density stored in the metal surface is strongly correlated on the ion emission~\cite{Colombier06}. A fast decaying ion signal that drops to zero in the first hundreds of femtosecond \cite{Schmidt00} was followed by a large peak of emission in the picosecond range \cite{Colombier06,Schmidt00,Stoian02}. 

\begin{figure}[htbp]
\begin{center}
\includegraphics[width=8.5cm]{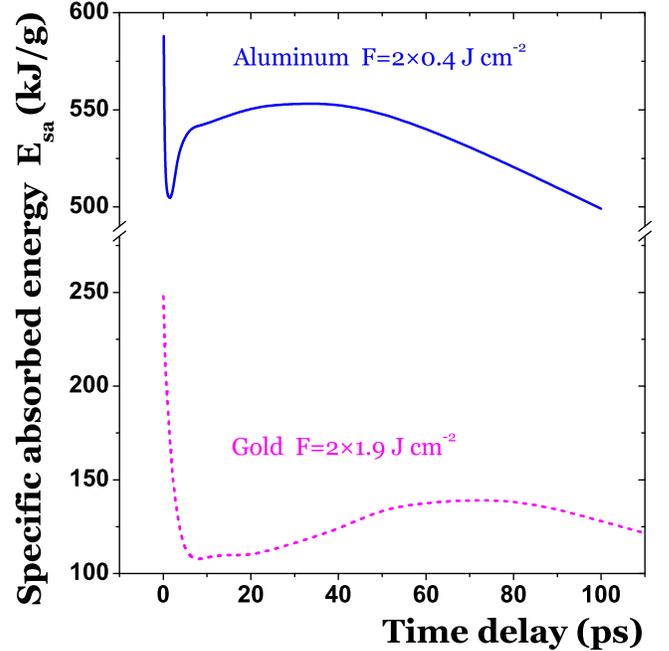}
\caption{Calculated specific absorbed energy in the skin depth of an aluminum target irradiated by a double pulse ($2 \times 0.4$ Jcm$^{-2}$) and a gold target irradiated by a double pulse ($2 \times 1.9$ Jcm$^{-2}$). Simulations were performed as a function of time delay.}\label{double}
\end{center}
\end{figure}

To discuss the experimental results related to the time varying ion signal, we have calculated the specific absorbed energy inside the skin depth in a metallic sample. We have chosen Al and Au as examples. The skin depth is defined as a time-dependent quantity, depending on the imaginary index which evolves with space and time. The laser irradiation is composed by two numerical identical pulses with a varying delay. We make the assumption that, to a certain extent, the energy density in the superficial layers determines the efficiency of phase transformation into a gas state, and, therefore, particle emission~\cite{Colombier06}. The total specific absorbed energy ($E_{sa}$) of a 800 nm, 150 fs double-pulse is then calculated as a function of time separation between the two pulses. We discuss two cases, Al and Au for two reasons. The first reason is related to the different values of \emph{(e-ph)} coupling, Al showing an almost 10 times higher electron-phonon interaction strength. Secondly, Al shows a particular optical characteristic, with no significant change between solid and liquid phases at 800 nm \cite{Krishnan93}. The numerical experiment are carried away with two identical sub-threshold pulses of 150 fs and 800 nm wavelength. The threshold is defined numerically by the onset of the gas-phase at 0.45 Jcm$^{-2}$ for Al and 2 Jcm$^{-2}$ for Au. The results are presented in Fig.~(\ref{double}). For both material, the following particularity were noted. 

Several characteristic stages are noticed with various peaks in the absorbed energy density. The first one appears during the first picosecond and shows a decay in the amount of energy stored $E_{sa}$, due to the decrease of $\Re_{e}[\tilde\sigma_{\beta}]$. This decay is faster for Al than for Au. Note that an important result not visible here is that simulations with a single pulse and a nonperturbing probe have revealed that the absorption increases strongly during the first pulse. This increase can not be observed with an identical double-pulse experiment due to a convoluting effect. This increase is a consequence of the instantaneous $T_{e}$ increase which yields a strong augmentation of the collision frequency. As $T_{e}$ reaches its maximum, the absorption is maximal and the observed minimum is almost synchronized with the corresponding time. The electron-phonon energy transfer is responsible for the $T_{e}$ decrease. As a consequence, $T_{e}$ can be maximal at the end of the first pulse if the electron-phonon coupling constant is very low. In the cases presented here, the absorption is maximal at about the half of the pulse duration, the corresponding time delay is close to the axis origin and a fast decay occurs for short delay times. This variation is related to the $T_{e}$ decrease which induces a drop of the collision frequency $\nu$ and a diminution of $\Re_{e}[\tilde\sigma_{\beta}]$. According to eq. (\ref{drude}) in the range $\omega_{\beta}/\nu>1$,  $\Re_{e}[\tilde\sigma_{\beta}]$ and the associated Joule heating process decreases as $\nu$ decreases. Moreover, a simulation performed, with setting $\nu_{neq}=0$, has shown that there is no decay for short time delay in this case. The electron-phonon energy transfer rate is correlated to the slope of the $E_{sa}$ curve at short times. This slope is steeper for Al than for Au and the decrease ends at 1.5 ps for Al and 6 ps for Au. A liquid phase is reached around 2 ps for Al and 7 ps for Au. The effect of the solid-liquid transition on the optical properties, which is expected to be more important for Au than for Al, is negligible because the nonequilibrium mechanism of absorption, especially \emph{(e-e)} umklapp  collisions, dominates. During this time, the ionic temperature $T_{i}$ increases and the $\nu^{ei}$ increase results in an augmentation of $\Re_{e}[\tilde\sigma_{\beta}]$, balancing the effect of $T_{e}$ decrease. At this stage, the optical properties are dominated by the ionic heating and $T_{i}$ drives the absorption mechanism. After a single pulse, the thermalisation is reached around 15 ps for Al and 50 ps for Au. $T_{i}$ begins to decrease while $\rho$ augments, prolonging the absorption increase. Note that solidification process begins to occur at the rear side of the liquid layer. At 35 ps for Al and 75 ps for Au, solidification reaches the free surface of the liquid layer and $\rho$ remains stable at solid density. The temperature decrease induces a drop of $E_{sa}$ for longer time delays.

These simulations have shown that a fast decay in the absorbed energy can be imputed to the electronic temperature decrease and associated \emph{(e-e)} effects during the nonequilibrium stage. Due to a smoother lattice heating, the fast decay is slower for Au than for Al. A recovery of the energy coupling efficiency on a tens picosecond scale is also observed for both materials. We have shown that the fact that the enhancement of optical absorption on picosecond scale is due to an elevation of the ionic temperature. This enhancement happens earlier for Al because of a high electron-phonon coupling constant. Recondensation of matter into a high density phase occurs after several tens of ps and result in a $E_{sa}$ drop. A similar situation was observed in experiments. As a conclusion, umklapp electron-electron collisions in the solid state determine the optical absorption behavior when the fluence is close to the threshold value.

\section{Conclusion}

Collisional absorption by inverse bremsstrahlung plays a major
role in laser-matter interactions. The full description of inverse
bremsstrahlung absorption requires knowledge of the
electron-photon absorption mechanism and the associated
umklapp electron-electron and electron-ion collision rates. We have conducted a simplified
modelling of the optical properties to perform realistic
simulations of several experiments present in the literature. We have used a relaxation-time model
 to approximate the two-body scattering rate. An important parameter
determining the modification of laser-light absorption is the
conductivity parameter and we have proposed a description of its evolution during
the material excitation. We used published results to test the hypothesis of an electronic contribution in an unified frame, covering various excitation stages, against its ability to explain experimental observations. In bringing together several accepted models in
solid and plasma range, a conductivity dependence of macroscopic
properties such as electronic temperature $(T_{e})$, ionic temperature
$(T_{i})$ and electronic density $(N_{e})$ was employed. This study is indicative to the potential role of electron-electron umklapp processes in optical absorption. An accurate evaluation of the umklapp rate would require complex kinetic approaches.

The numerical diagnostic that we have developped provide the correlation between
the sequence of states of matter and the optical response. The information, resolved in time, give insights into the relaxation mechanisms which drive the phase transitions and open new opportunities to monitor
the excitation of a metal in controling both its optical and thermal evolution. Several regimes of irradiation, including above and subthreshold regimes were investigated. Clear modification of electronic and hydrodynamic effects was observed. Beyond the retrieval of the optical properties and the discussion of associated thermodynamic states, a qualitative idea on the transient energy coupling properties was given. 

\section*{ACKNOWLEDGEMENTS}
The authors acknowledge the support of the GIP ANR and PICS programs.

\end{document}